\documentclass[aps,prb,superscriptaddress,twocolumn,showpacs]{revtex4}

\usepackage{hyperref}
\usepackage{epsfig}
\usepackage{hyperref}
\usepackage{epsfig}

\newcommand{\mb}[1]{ { \mbox{\boldmath{$#1$}}}  } 
\newcommand{\fig}[1]{Fig.~\ref{#1}}

\newcommand{\expect}[1]{\langle #1 \rangle}

\begin{document}
 
\title{Constructive influence of the induced electron pairing on the Kondo state}

\author{T. Doma\'nski}
\email{doman@kft.umcs.lublin.pl}
\affiliation{Institute of Physics, M.\ Curie Sk\l odowska University, 20-031 Lublin, Poland}

\author{I. Weymann}
\email{weymann@amu.edu.pl}
\affiliation{Faculty of Physics, A.\ Mickiewicz University, 
			 ul. Umultowska 85, 61-614 Pozna{\'n}, Poland}

\author{M. Bara\'nska}
\affiliation{Institute of Physics, Polish Academy of Sciences, 02-668 Warsaw, Poland} 
\affiliation{Institute of Physics, M.\ Curie Sk\l odowska University, 20-031 Lublin, Poland}

\author{G. G\'orski}
\affiliation{Faculty of Mathematics and Natural Sciences, University of Rzesz\'ow, 
       35-310 Rzesz\'ow, Poland}

      \date{\today}

\begin{abstract}
Superconductivity and magnetism are usually the conflicting
(competing) phenomena. We show, however, that in nanoscopic objects
the electron pairing may promote the magnetic ordering. 
Such situation is possible at low temperatures in the quantum dots 
placed between the conducting and superconducting reservoirs, where  
the proximity induced electron pairing cooperates with the correlations 
enhancing the spin-exchange interactions. The emerging Kondo resonance,
which is observable in the Andreev conductance, can be significantly 
enhanced by  the coupling to superconducting lead. We explain 
this intriguing tendency within the Anderson impurity model using: 
the generalized Schrieffer-Wolff canonical transformation, the second 
order perturbative treatment of the Coulomb repulsion, and the 
nonperturbative numerical renormalization group calculations.
We also provide  hints for experimental observability of this phenomenon. 
\end{abstract}  

\pacs{73.23.-b,73.21.La,72.15.Qm,74.45.+c}


\maketitle

\section*{Introduction}

\begin{figure}
\includegraphics[width=0.6\columnwidth]{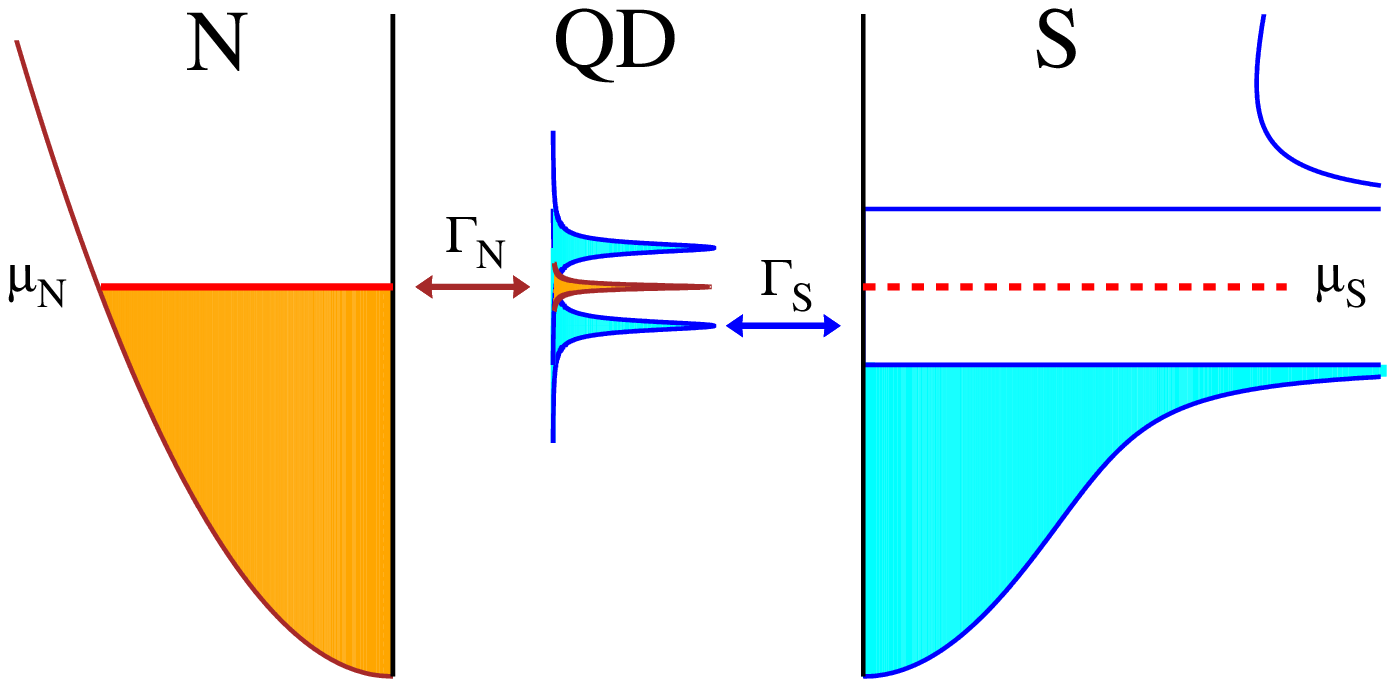}
\includegraphics[width=0.6\columnwidth]{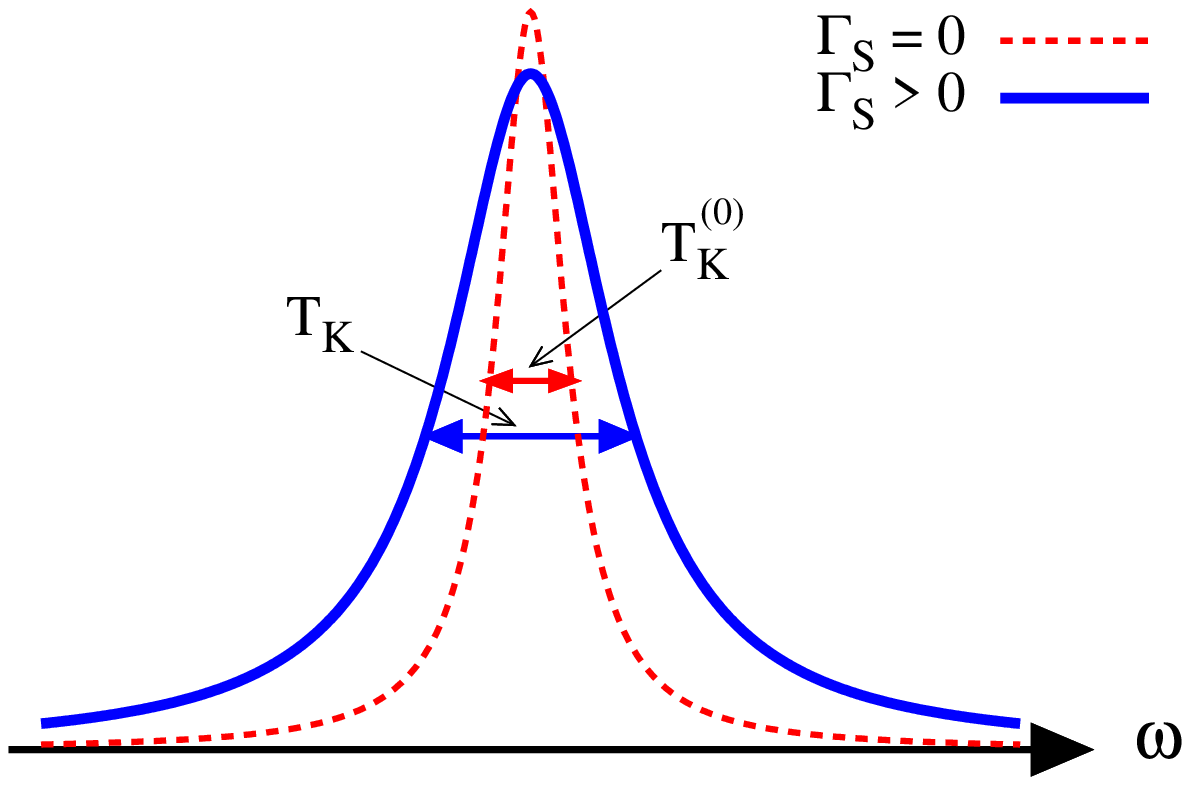}
\caption{(Color online)
Schematic view of the energy spectrum of the N-QD-S junction in the spinfull 
doublet configuration (top panel), where the QD Andreev bound states 
(driven by the coupling $\Gamma_{S}$ to superconducting reservoir) coexist 
with the zero-energy Kondo peak (originating from the Coulomb potential 
$U_{d}$ and the coupling $\Gamma_{N}$ to metallic lead). Bottom panel 
shows, that width and height of the Kondo resonance strongly depend 
on $\Gamma_{S}$.}
\label{scheme_setup}
\end{figure}

Correlated quantum impurity immersed in the Fermi sea usually develops 
the  spin-exchange interactions \cite{S-W}, that cause its total 
(or partial) screening below some characteristic (Kondo) temperature 
$T_{K}$  \cite{Kondo,hewson_book}. This effect is manifested in the 
quantum impurity spectrum by the Abrikosov-Suhl peak appearing 
at the Fermi level. It has been predicted \cite{GlazmanJETP88,NgPRL88} 
and experimentally confirmed \cite{goldhaber-gordon_98,cronenwett_98} 
that in the correlated quantum dot (QD) embedded between  metallic 
electrodes, such effect enhances the zero-bias tunneling conductance 
\cite{Pustilnik-04}. This situation changes, however, if one (or both) 
external electrode(s) is (are) superconducting because of the proximity 
induced electron pairing \cite{Rodero-11,Viewpoint-13}. 
Depending on the energy level $\varepsilon_{d}$, Coulomb potential $U_{d}$ 
and the coupling $\Gamma_{S}$  to superconducting reservoir, the ground 
state may evolve from the spinful configuration $\left| \sigma \right>$ 
(where $\sigma=\uparrow,\downarrow$) to the spinless BCS-type state $u_{d} 
\left| 0 \right> - v_{d} \left| \uparrow \downarrow \right>$ \cite{Bauer-07}. 
Such quantum phase transition (QPT) has a qualitative influence on the 
spin-screening mechanism \cite{Viewpoint-13}. In this work we show 
that, for $\Gamma_{S} \leq U_{d}$, the proximity induced electron pairing 
can strongly amplify the screening effects and give rise to a broadening  
of the Kondo peak \cite{Weymann-14,Zitko-15} (see fig.\ \ref{scheme_setup}). 

At first glance, such tendency seems to be rather counter-intuitive 
because $\Gamma_S$ supports the proximity induced electron pairing 
that should compete with the magnetism.  We provide microscopic 
arguments explaining this intriguing result, based on three 
independent methods. Our study might stimulate and guide future 
experimental attempts to verify this theoretical prediction 
in the N-QD-S heterostructures (schematically displayed in the left 
panel of fig.\ \ref{scheme_setup}), using e.g. 
self-assembled InAs quantum islands \cite{Deacon-10}, 
semiconducting quantum wires \cite{Lee-12,Aguado-13} 
or carbon nanotubes \cite{Pillet-13,Schindele-13}. Former 
measurements of the subgap differential conductance have 
already provided  evidence for the Andreev/Shiba bound states 
\cite{Andreev,Yu-Shiba-Rusinov,BalatskyRMP06} and a tiny 
(but clear) signature of the zero-bias anomaly driven by the 
Kondo effect \cite{Deacon-10,Aguado-13,Lohneysen-12,Chang-13}. 
Its variation with respect to the  ratio $\Gamma_{S}/U_{d}$ 
has not been investigated carefully enough, but this seems 
to be feasible.

Similar zero-bias anomalies driven by the superconducting proximity 
effect are nowadays intensively explored also in the quantum wires 
coupled to the $s$-wave superconductors, signaling the Majorana-type 
quasiparticles \cite{Mourik-12,Yazdani-14,Franke-15}.  These exotic 
quasiparticles originate solely from the Andreev/Shiba states in 
the presence of the strong spin-orbit interaction and the Zeeman 
effect \cite{Chevallier-13}. The present study might hence be 
useful for distinguishing the zero-bias enhancement due to the 
Kondo effect from the one driven by the Majorana-type quasiparticles.

\section*{Results}

In what follows we address the proximity induced electron pairing and 
study its feedback on the Kondo state, focusing on the deep subgap regime. 
First, we introduce the model and discuss its simplified version relevant 
for the deep subgap states. Next, we discuss the issue of singlet-doublet 
quantum phase transition in the limit of negligible coupling to the normal lead,
$\Gamma_{N}\rightarrow 0$, 
emphasizing its implications for the Kondo-type correlations. 
We then determine the effective spin exchange potential, generalizing 
the Schrieffer-Wolff transformation \cite{S-W} for the proximized 
quantum dot, and confront the estimated Kondo temperature with 
the nonperturbative NRG data (showing excellent quantitative 
agreement over the region $\Gamma_{S}\leq 0.9U_{d}$). 
We also  discuss the results obtained from the second-order 
perturbation theory (SOPT) with respect to the Coulomb potential, 
that provide an independent evidence for the Kondo temperature 
enhancement by increasing $\Gamma_{S}$ (in the doublet state). 
Finally, we discuss the experimentally measurable conductance 
for the subgap regime and give a summary of our results.

\subsection*{Microscopic model in the subgap regime}

For the description of the N-QD-S junction we use 
the Anderson impurity model\cite{Anderson}
\begin{eqnarray} 
\hat{H} &=& \sum_{\beta}\hat{H}_{\beta} + \sum_{\sigma}  
\varepsilon_{d} \hat{d}^{\dagger}_{\sigma} \hat{d}_{\sigma}  
+  U_{d} \; \hat{n}_{d \uparrow} \hat{n}_{d \downarrow}  
\nonumber \\ &+&
\sum_{{\bf k},\sigma}\sum_{{\beta}}  \left( V_{{\bf k} \beta} \; 
\hat{d}_{\sigma}^{\dagger}  \hat{c}_{{\bf k} \sigma \beta } 
+  V_{{\bf k} \beta}^{*}  \hat{c}_{{\bf k} \sigma 
\beta }^{\dagger} \hat{d}_{\sigma} \right) ,
\label{model}
\end{eqnarray} 
where $\beta$ refers to the normal ($\beta=N$) and superconducting 
($\beta=S$) electrodes, respectively.
The operator $\hat{d}_\sigma^{(\dag)}$ annihilates (creates)
an electron with spin $\sigma$ and
energy $\varepsilon_{d}$ in the quantum dot, while
$V_{{\bf k} \beta}$ denotes the tunneling matrix elements.
The repulsive Coulomb potential is denoted by $U_d$
and $\hat{n}_{d \sigma} = \hat{d}_{\sigma}^\dag \hat{d}_{\sigma}$.
Itinerant electrons of the metallic 
reservoir are treated as free fermions,
$\hat{H}_{N} \!=\! \sum_{{\bf k},\sigma} \xi_{{\bf k}N}  
\hat{c}_{{\bf k} \sigma N}^{\dagger} \hat{c}_{{\bf k} \sigma N}$,  
and the isotropic superconductor is described by the BCS model
$\hat{H}_{S} \!=\!\sum_{{\bf k},\sigma}  \xi_{{\bf k}S}
\hat{c}_{{\bf k} \sigma S }^{\dagger}  \hat{c}_{{\bf k} \sigma S} 
\!-\! \sum_{\bf k} \Delta  \left( \hat{c}_{{\bf k} \uparrow S }
^{\dagger} \hat{c}_{-{\bf k} \downarrow S }^{\dagger} + \hat{c}
_{-{\bf k} \downarrow S} \hat{c}_{{\bf k} \uparrow S }\right)$.
Here, $\hat{c}_{\mathbf{k}\sigma\beta}^{(\dag)}$
denotes the annihilation (creation) operator of a spin-$\sigma$
electron with momentum $\mathbf{k}$ of energy 
$\xi_{{\bf k}\beta}$ in the lead $\beta$,
while $\Delta$ denotes the superconducting energy gap.
It is convenient to introduce the characteristic couplings 
$\Gamma_{\beta}=2\pi \sum_{\bf k} |V_{{\bf k}\beta}|^2 \;  
\delta(\omega \!-\! \xi_{{\bf k}\beta})$, assuming that they 
are constant within the subgap energy regime $|\omega| \leq \Delta$. 

Since we are interested in a relationship between the Andreev/Shiba 
quasiparticles and the Kondo state we can simplify the considerations 
by restricting ourselves to an equivalent Hamiltonian \cite{rozhkov}
\begin{eqnarray} 
\hat{H} &=&  \hat{H}_{N} + \sum_{{\bf k},\sigma}  \left( V_{{\bf k} N} \; 
\hat{d}_{\sigma}^{\dagger}  \hat{c}_{{\bf k} \sigma N} 
+  V_{{\bf k} N}^{*}  \hat{c}_{{\bf k} \sigma 
N}^{\dagger} \hat{d}_{\sigma} \right)  
\nonumber \\& + & \sum_{\sigma}  
\varepsilon_{d} \hat{d}^{\dagger}_{\sigma} \hat{d}_{\sigma}  
+  U_{d} \; \hat{n}_{d \uparrow} \hat{n}_{d \downarrow}  
- \left( \Delta_{d} \hat{d}_{\uparrow}^{\dagger} 
\hat{d}_{\downarrow}^{\dagger} + \mbox{\rm h.c.} \right) 
\label{proximized} 
\end{eqnarray} 
relevant for the subgap regime in a weak coupling limit 
$\Gamma_{S} < \Delta$. Effects due to the superconducting 
electrode are here played by the induced on-dot pairing gap 
$\Delta_{d}=\Gamma_{S}/2$ \cite{Rodero-11,Bauer-07}. This Hamiltonian 
(\ref{proximized}) neglects the high-energy states existing outside 
the energy gap window $|\omega| \geq \Delta$ (see Methods
for a discussion) that are irrelevant for the present context.

\subsection*{Subgap quasiparticles of the proximized quantum dot}

To understand the influence of electron pairing on the Kondo
effect, it is useful to recall basic aspects of the singlet-doublet 
quantum phase transition in the 'superconducting atomic limit' 
$\Gamma_{N}\rightarrow 0$ \cite{Rodero-11,Tanaka-07}. Exact eigenstates of 
the proximized QD are then represented  either by the spinfull configurations 
$\left| \sigma \right>$ with eigenenergy $\varepsilon_{d}$, or the spinless 
(BCS-type) states 
\begin{eqnarray}
\left| {-} \right> & = & u_{d} \left| 0 \right> - v_{d} \left|
\uparrow \downarrow \right> ,\\
\left| {+} \right> & = & v_{d} \left| 0 \right> + u_{d} \left|
\uparrow \downarrow \right> ,
\end{eqnarray}
whose eigenvalues are
\begin{eqnarray}
E_{\mp}=\left( \varepsilon_{d}+\frac{U_{d}}{2} \right) \mp
\sqrt{\left( \varepsilon_{d}+\frac{U_{d}}{2} \right)^{2}+\Delta_{d}^{2}} \,,
\label{qp_energies}
\end{eqnarray}
with the BCS coefficients 
\begin{eqnarray}
u_{d}^{2} = \frac{1}{2} \left[ 1 + \frac{\varepsilon_{d}+U_{d}/2}
{\sqrt{\left(\varepsilon_{d}+U_{d}/2\right)^{2}+\Delta_{d}^{2}}} 
\right] =1 -v_{d}^{2} .
\label{BCS_coef}
\end{eqnarray}
The single particle excitations,  between the doublet and 
singlet configurations, give rise to the following quasiparticle branches 
$\pm U_{d}/2\pm E_{d}$, where $E_{d}=\sqrt{(\varepsilon_{d}+U_{d}/2)^{2}
+(\Gamma_{S}/2)^{2}}$. Two energies $\pm \left( U_{d}/2- E_{d}\right)$ 
can be regarded as the low-energy excitations, whereas the other ones 
(shifted from them by $U_{d}$) represent the high-energy features. In 
realistic systems (where $U_{d}$ is typically much larger than $\Delta$) 
the latter ones usually coincide with a continuum formed outside the subgap regime
\cite{Deacon-10,Lee-12,Aguado-13,Pillet-13,Schindele-13,FranceschiNatNano2010}.

\begin{figure}
\includegraphics[width=1\columnwidth]{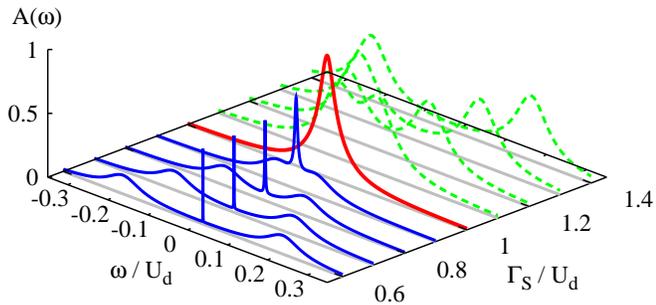}
\caption{(Color online)
The normalized spectral function $A(\omega)=\frac{\pi}{2}\Gamma_{N}
\rho_{d}(\omega)$ of the half-filled quantum dot obtained from 
the superconducting atomic limit solution (using the quasiparticle 
broadening $\Gamma_{N} = 10^{-1}\Gamma_{S}$) superposed with the 
Abrikosov-Suhl peak whose width $T_{K}$ is expressed by
Eq. (\ref{T_K}). The solid/dashed lines correspond to the doublet/singlet 
ground state configuration and the thick-red curve indicates 
the quantum phase transition at $\Gamma_{S}\!=\!U_{d}$.}
\label{figure2}
\end{figure}

Diagonal part of the single particle Green's function (for its definition 
see Methods) is in the subgap regime  given by \cite{Bauer-07}
\begin{eqnarray}
{\mb G}_{11}(\omega)
& = &   \frac{\alpha \;\; u_{d}^{2}}
{\omega+\left( \frac{U_{d}}{2}-E_{d} \right)}+ \frac{ \alpha 
\;\; v_{d}^{2}}{\omega-\left( \frac{U_{d}}{2}-E_{d} \right)}
 \label{G11_atomic} \,, 
\end{eqnarray}
where the partial spectral weight is
$\alpha = \left[\mbox{\rm exp}\left( {\frac{U_{d}}{2T}}\right)
\right.$ $\left.
+\mbox{\rm exp}\left(\frac{E_{d}}{T}\right) 
\right]/\left[ 2\;\mbox{\rm exp}\left( 
{\frac{U_{d}}{2T}}\right)+\mbox{\rm exp}\left( 
\frac{-E_{d}}{T}\right) +\mbox{\rm exp}\left( 
\frac{E_{d}}{T}\right)\right]$
and we set the Boltzmann constant
equal to unity, $k_B\equiv 1$. The missing amount of 
the spectral weight $1-\alpha$ belongs to the high-energy 
states existing outside the superconductor gap.
At zero temperature, the subgap weight changes abruptly from 
$\alpha=0.5$ (when $E_{d}<U_{d}/2$) to $\alpha=1$ (when $E_{d}>U_{d}/2$).
At $E_{d}=U_{d}/2$ the quasiparticle crossing is a signature 
of the quantum phase transition from the doublet $\left| \sigma \right>$ to 
the singlet configuration $\left| {-} \right>$
\cite{Bauer-07,Rodero-11,Zitko-15}.

For infinitesimally small coupling $\Gamma_{N}$ one can extend the atomic limit 
solution (\ref{G11_atomic}) by imposing the quasiparticle broadening  
${\mb G}(\omega) \rightarrow {\mb G}(\omega+\frac{i}{2}\Gamma_{N})$. 
Figure \ref{figure2} shows the normalized spectral function
$A(\omega)=\frac{\pi}{2}\Gamma_{N} \rho_{d}(\omega)$,
with $\rho_{d}(\omega) \equiv -\pi^{-1} \mbox{\rm Im} {\mb G}_{11}(\omega)$,
for the half-filled quantum dot, $\varepsilon_d = -U_d/2$. On top of these
curves we have added the Abrikosov-Suhl peak (at $\omega=0$) whose 
width is given by the Kondo temperature, see Eq. (\ref{T_K}).
Upon increasing the ratio $\Gamma_{S}/U_{d}$,
the Andreev quasiparticle peaks  move closer and 
they ultimately merge at the critical point $\Gamma_{S}=U_{d}$, 
and simultaneously the Abrikosov-Suhl peak gradually broadens 
all the way up to the QPT. For $\Gamma_{S}>U_{d}$, the Andreev 
peaks drift away from each other (see the dashed lines in Fig. 
\ref{figure2}) and the Kondo feature disappears for
the reasons discussed in the next subsection.

\subsection*{Spin exchange interactions and Kondo temperature}

Adopting the Schrieffer and Wolff approach \cite{S-W} to 
the Hamiltonian (\ref{proximized}) of the proximized quantum dot 
we can design  the canonical transformation 
\begin{eqnarray}
\hat{\tilde{H}} = e^{\hat{S}} \hat{H} e^{-\hat{S}} ,
\label{can_trans}
\end{eqnarray}
which perturbatively eliminates the hybridization term $\hat{V}=\sum_{{\bf k},\sigma}  
\left(V_{{\bf k} N} \; \hat{d}_{\sigma}^{\dagger}  \hat{c}_{{\bf k} \sigma N} 
+  V_{{\bf k} N}^{*}  \hat{c}_{{\bf k} \sigma N}^{\dagger} \hat{d}_{\sigma} \right)$.
To simplify the notation, we skip the subindex $N$ that unambiguously refers to 
the metallic lead. The terms linear in $V_{{\bf k} N}$ can be cancelled  
in the transformed Hamiltonian $\hat{\tilde{H}}$ by
choosing the operator $\hat{S}$ from the following constraint
\begin{eqnarray}
\left[ \hat{H}_{0},\hat{S} \right] = \hat{V} ,
\label{single_step}
\end{eqnarray}
where $\hat{H}_{0}=\hat{H}-\hat{V}$. For the Hamiltonian 
(\ref{proximized}) this can be satisfied with the anti-hermitian 
operator $\hat{S}=\hat{S}_{0} - \hat{S}^{\dagger}_{0}$, where 
\begin{eqnarray}
\hat{S}_{0}&=&\sum_{{\bf k}}\sum_{\alpha=+,-} \gamma_{1,{\bf k}}^{\alpha} 
\left( \hat{c}_{{\bf k}\uparrow}^{\dagger}\hat{d}_{\uparrow} 
\hat{n}_{d\downarrow}^{\alpha} + \hat{c}_{{\bf k}\downarrow}^{\dagger}
\hat{d}_{\downarrow} \hat{n}_{d\uparrow}^{\alpha} \right) 
\nonumber \\ & - &
 \sum_{\bf k} \sum_{\alpha=+,-} \gamma_{2,{\bf k}}^{\alpha} \left(
\hat{c}_{{\bf k}\uparrow}^{\dagger}\hat{d}_{\downarrow}^{\dagger} 
\hat{n}_{d\uparrow}^{\alpha} -
\hat{c}_{{\bf k}\downarrow}^{\dagger}\hat{d}_{\uparrow}^{\dagger} 
\hat{n}_{d\downarrow}^{\alpha} \right) 
\label{generator}
\end{eqnarray}
and \cite{S-W}
\begin{eqnarray}
\hat{n}_{d\sigma}^{\alpha} =
\left\{ 
\begin{array}{ll}
\hat{d}_{\sigma}^{\dagger} \hat{d}_{\sigma} & \mbox{\rm for} \hspace{0.1cm} 
\alpha=+ ,\\ 1- \hat{d}_{\sigma}^{\dagger} \hat{d}_{\sigma} & \mbox{\rm for} 
\hspace{0.1cm} \alpha=- .  
\end{array}
\right.
\end{eqnarray}
The second term of Eq. (\ref{generator}) explicitly differs from the standard 
operator used by Schrieffer and Wolff \cite{S-W}.
From the lengthy by straightforward algebra we find that
the constraint (\ref{single_step}) implies the following coefficients 
$\gamma_{\nu,{\bf k}}^{\alpha}$  
\begin{eqnarray}
\gamma_{1,{\bf k}}^{+}&=&\frac{(\xi_{\bf k}+\varepsilon_{d})V_{\bf k}}
{\xi_{\bf k}^2 - U_{d}(\xi_{\bf k}+\varepsilon_{d})-(\varepsilon_{d}^2
+\Delta_{d}^2)} , \label{coef1} \\
\gamma_{2,{\bf k}}^{+}&=&\frac{\Delta_{d}V_{\bf k}}{\xi_{\bf k}^2 
+ U_{d}(\xi_{\bf k}-\varepsilon_{d})-(\varepsilon_{d}^2+\Delta_{d}^2)},
\label{coef3_4} \\
\gamma_{1,{\bf k}}^{-}&=&\frac{V_{\bf k}}{\xi_{\bf k}-\varepsilon_{d}}
-\frac{\Delta_{d}}{\xi_{\bf k}-\varepsilon_{d}} \gamma_{\bf 2,k}^{+} , 
\label{coef2} \\
\gamma_{2,{\bf k}}^{-}&=&\frac{\Delta_{d}}{\xi_{\bf k}
+\varepsilon_{d}}\gamma_{\bf 1,k}^{+}.
\label{coef3_5}
\end{eqnarray}
For $\Delta_{d}=0$, the coefficients $\gamma_{2,{\bf k}}^{\alpha}$ 
identically vanish and the other ones, given by Eqs.~(\ref{coef1},\ref{coef2}),
simplify to the standard expressions  $\gamma_{1,{\bf k}}^{+}=V_{\bf k}/\left( 
\xi_{\bf k}-U_{d}-\varepsilon_{d}\right)$ and $\gamma_{1,{\bf k}}^{-} 
= V_{\bf k}/\left(\xi_{\bf k}-\varepsilon_{d}\right)$ 
of the Schrieffer-Wolff transformation. \cite{S-W}

In the transformed Hamiltonian 
\begin{widetext}
\begin{eqnarray}
\hat{{\tilde H}} &=& \sum_{\bf k \sigma} \xi_{\bf k} 
\hat{c}_{\bf k \sigma}^{\dagger} \hat{c}_{\bf k \sigma} 
+ \frac{1}{2}  \sum_{\bf k p \sigma} \gamma_{1,{\bf  k}}^{-} V_{\bf p}
\left( \hat{c}_{\bf k \sigma}^{\dagger} \hat{c}_{\bf p \sigma} + \mbox{\rm h.c.} \right)
+ \sum_{\sigma} \left( \varepsilon_{d} - \sum_{\bf k} \gamma_{1,{\bf  k}}^{-} 
V_{\bf k} \right ) \hat{d}_{\sigma}^{\dagger} \hat{d}_{\sigma} \nonumber \\
&-& \left ( \Delta_{d} + \sum_{\bf k} \gamma_{2,{\bf k}}^{-}V_{\bf k} \right)
\left (\hat{d}_{\uparrow}^{\dagger} \hat{d}_{\downarrow}^{\dagger} +
\mbox{\rm h.c.} \right) + \left[ U_{d} + 2 \sum_{\bf k} 
 \left(\gamma_{1,{\bf  k}}^{-} - \gamma_{1,{\bf  k}}^{+} \right) V_{\bf k} \right] 
\hat{n}_{d\uparrow} \hat{n}_{d\downarrow} \nonumber \\
&-& \frac {1}{2} \sum_{\bf k p} \gamma_{2,{\bf  k}}^{-}
V_{\bf p} \left [ \left( \hat{c}_{{\bf k}\uparrow}^{\dagger}
\hat{c}_{{\bf p}\downarrow}^{\dagger}+ \hat{c}_{{\bf p}\uparrow}^{\dagger}
\hat{c}_{{\bf k}\downarrow}^{\dagger} \right) + \mbox{\rm h.c.} \right ] 
+ \frac{1}{2} \sum_{\bf k p \sigma} 
\left ( \gamma_{1,{\bf  k}}^{-} - \gamma_{1,{\bf  k}}^{+} \right)
V_{\bf p} \left ( \hat{c}_{{\bf k}\sigma}^{\dagger}
\hat{c}_{{\bf p}-\sigma}^{\dagger} \hat{d}_{-\sigma} \hat{d}_{\sigma} 
+ \mbox{\rm h.c.} \right) \nonumber \\
&+&\frac{1}{2} \sum_{\bf k p} \left( \gamma_{2,{\bf  k}}^{-} 
-\gamma_{2,{\bf  k}}^{+} \right) 
V_{\bf p} \left( {\hat c}_{\bf k \uparrow}^{\dagger}
{\hat c}_{\bf p \downarrow}^{\dagger}{\hat d}_{\uparrow}^{\dagger} 
{\hat d}_{\bf \uparrow} + {\hat c}_{\bf p \uparrow}^{\dagger}
{\hat  c}_{\bf k \downarrow}^{\dagger}{\hat d}_{\bf \downarrow}^{\dagger} 
{\hat d}_{\bf \downarrow} + {\hat d}_{\bf \uparrow}^{\dagger}
{\hat d}_{\bf \downarrow}^{\dagger}{\hat c}_{\bf k \uparrow}^{\dagger}
{\hat c}_{\bf p \uparrow}+{\hat d}_{\bf \uparrow}^{\dagger}
{\hat d}_{\bf \downarrow}^{\dagger}{\hat c}_{\bf k \downarrow}^{\dagger}  
{\hat c}_{\bf p \downarrow} + \mbox{\rm h.c.} \right ) \nonumber \\
&+&\frac{1}{2} \sum_{\bf k p \sigma} \left( \gamma_{1,{\bf  k}}^{+} 
- \gamma_{1,{\bf  k}}^{-} \right) V_{\bf p} 
\left[ \left ( {\hat c}_{\bf k \sigma}^{\dagger}
{\hat c}_{\bf p \sigma}{\hat d}_{-\sigma}^{\dagger} 
{\hat d}_{-\sigma} + {\hat c}_{\bf k \sigma}^{\dagger}
{\hat  d}_{\sigma}{\hat d}_{-\sigma}^{\dagger} 
{\hat c}_{\bf p -\sigma} \right ) + \mbox{\rm h.c.} \right ]
\label{H_jj_flow}
\end{eqnarray}
\end{widetext}
we can recognize: the spin exchange 
term, the interaction between QD and itinerant electrons, 
the pair hopping term, and renormalization of the QD energy and 
the on-dot pairing. Since we focus on the screening effects, 
we study in detail only the effective spin-exchange term
\begin{eqnarray}
\hat{H}_{exch} = -  \sum_{{\bf k}, {\bf p}} \; J_{\bf k p} \;\;
\hat{\bf S}_{d} \cdot \hat{\bf S}_{{\bf k}{\bf p}} , 
\end{eqnarray}
where $\hat{\bf S}_{d}$ describes the spin operator of the dot and 
$\hat{\bf S}_{{\bf k}{\bf p}}$ refers to the spins of itinerant 
electrons in metallic lead. 
Other contributions are  irrelevant for the Kondo physics.

Formal expression for the effective exchange potential 
\begin{eqnarray}
J_{{\bf k},{\bf p}}= \frac{1}{2}\left[ \left( \gamma_{1,{\bf k}}^{+}
- \gamma_{1,{\bf k}}^{-} \right) V_{\bf p} +
\left( \gamma_{1,{\bf p}}^{+}
- \gamma_{1,{\bf p}}^{-} \right) V_{\bf k} \right]
\label{S-W-like}
\end{eqnarray}
is analogous to the standard Schrieffer-Wolff result, \cite{S-W}
but here we have different coefficients
$\gamma_{1,{\bf k}}^{\pm}$ expressed in Eqs.~(\ref{coef1},\ref{coef2}).
This important aspect generalizes the Schrieffer-Wolf potential \cite{S-W}
and captures the effects induced by the on-dot pairing. 

In particular, near the Fermi momentum the exchange potential (\ref{S-W-like}) 
simplifies to
\begin{eqnarray}
J_{{\bf k}_{F},{\bf k}_{F}}=\frac{ U_{d} \left| V_{{\bf k}_{F}}\right|^{2}}
{\varepsilon_{d} \left( U_{d} + \varepsilon_{d} \right)+\Delta_{d}^{2}} . 
\label{generalized_SW}
\end{eqnarray}
It is worthwhile to emphasize that this formula (\ref{generalized_SW}) precisely 
reproduces constraint for the quantum phase transition discussed in the previous section.
To prove it, we remark that $J_{{\bf k}_{F},{\bf k}_{F}}$ changes 
discontinuously from the negative (antiferromagnetic) to the positive  
(ferromagnetic) values at $\varepsilon_{d} \left( U_{d} + \varepsilon_{d} 
\right)+\Delta_{d}^{2}=0$.
Such changeover occurs thus at
\begin{eqnarray}
 \left( \varepsilon_{d} + \frac{U_{d}}{2}\right)^{2} + \left( 
\frac{\Gamma_{S}}{2} \right)^{2} = \left( \frac{U_{d}}{2}\right)^{2}  \,,
\label{ala_Bauer}
\end{eqnarray}
which is identical to the  QPT constraint 
$E_{d}^{2}=(U_{d}/2)^{2}$ originally derived 
in Ref.~\cite{Bauer-07}.

\begin{figure}
\includegraphics[width=0.8\columnwidth]{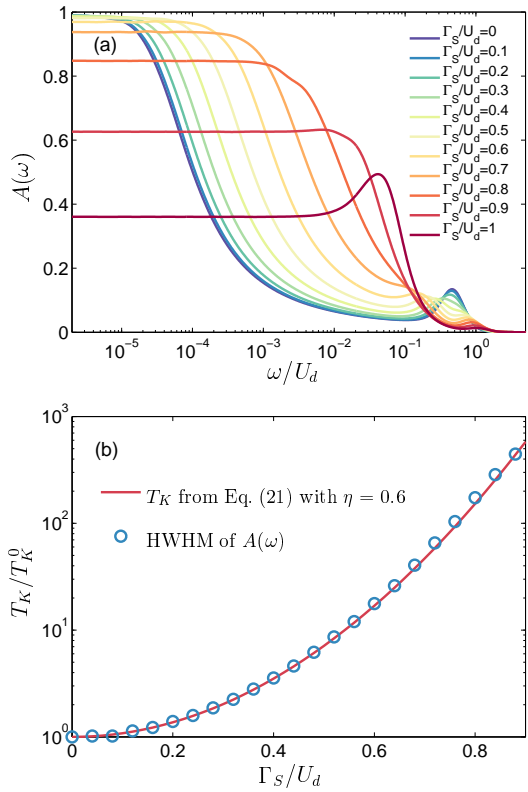}
\caption{
(Color online)
(a) The normalized spectral function $A(\omega)$
of the correlated quantum dot obtained by NRG 
for the model Hamiltonian (\ref{proximized}) at half-filling
for different values of $\Gamma_S$, as indicated.
Note the logarithmic energy scale in panel (a).
(b) The Kondo temperature $T_K$ extracted from
the half width at half maximum (HWHM) of $A(\omega)$
from panel (a) (circles)
and $T_K$ obtained from Eq.~(\ref{T_K}) with $\eta = 0.6$ (solid line).
$T_K^0$ denotes the Kondo temperature in the case of $\Gamma_S = 0$,
$T_K^0 \approx 10^{-5}$.
The parameters are: $U_d=0.1$ and $\Gamma_N = 0.01$.
All parameters are in units of band halfwidth $W\equiv 1$.
}
\label{TK_NRG}
\end{figure}

To estimate the effective Kondo temperature in the case of spinfull 
configuration (for $\Gamma_S < U_d$), we use the formula,
\cite{Haldane-78,Tsvelik-83} $T_{K} = \frac{2}{\pi} D \; \mbox{\rm exp}
\left\{ -\phi \left [ 2\rho(\varepsilon_{\bf F}) J_{{\bf k}_F 
{\bf k}_F} \right ]\right\}$, where $\rho(\varepsilon_{F}$) is 
the density of states at the Fermi level, $D$ is the cut-off energy
and the auxiliary function is defined as $\phi(y) \simeq |y|^{-1} - 0.5 \ln{|y|}$.
In  present case the Kondo temperature is expressed by
\begin{eqnarray}
T_{K} = \eta \, \frac{\sqrt{\Gamma_N U_d}}{2} \; {\rm exp} \!\left[ \frac{ \varepsilon_{d} \left( 
\varepsilon_{d} +U_{d} \right)+\Delta_{d}^{2} }{\Gamma_N U_d/\pi}\right] ,
\label{T_K}
\end{eqnarray}
with $\eta$ being a constant of the order of unity.
Influence of the on-dot pairing on the Kondo temperature can be well illustrated
considering  the half-filled quantum dot case $\varepsilon_{d}\!=\!-U_{d}/2$. 
The spin exchange potential (\ref{generalized_SW}) is then given by 
\begin{eqnarray}
J_{{\bf k}_{F},{\bf k}_{F}}= \frac{ - \; 4 U_{d} \left| V_{{\bf k}_{F}}
\right|^{2}} { U_{d}^{2} - \left( 2 \Delta_{d} \right)^{2}} 
\label{suppl_18}
\end{eqnarray}
and for $\Delta_{d}=0$ it reproduces the standard Schrieffer-Wolff result \cite{S-W} 
\begin{eqnarray}
J_{{\bf k}_{F},{\bf k}_{F}}^{N} = - \; \frac{  4  \left| V_{{\bf k}_{F}}
\right|^{2}}{U_{d}}  
\label{normal}
\end{eqnarray}
characteristic for the impurity hosted in the metallic 
reservoir. The relative change of $J_{{\bf k}_{F},{\bf k}_{F}}$
arising from the on-dot pairing is 
\begin{eqnarray}
\frac{J_{{\bf k}_{F},{\bf k}_{F}}}{J_{{\bf k}_{F},{\bf k}_{F}}^{N}}
= \frac{U_{d}^{2}}{U_{d}^{2} - \left( 2 \Delta_{d} \right)^{2}} 
= \frac{1}{1-\left( \Gamma_S/U_{d}\right)^{2}} .
\label{ratio}
\end{eqnarray}
For the doublet phase ($\Gamma_S<U_d$) the exchange coupling is 
antiferromagnetic, whereas for the singlet state ($\Gamma_S>U_d$) 
it becomes ferromagnetic. In the latter case, however, such 
ferromagnetic interactions are ineffective because the spinless BCS singlet, 
$u_{d} \left| 0 \right> - v_{d} \left| \downarrow\uparrow \right>$,
cannot be screened.

\begin{figure}
\includegraphics[width=1\columnwidth]{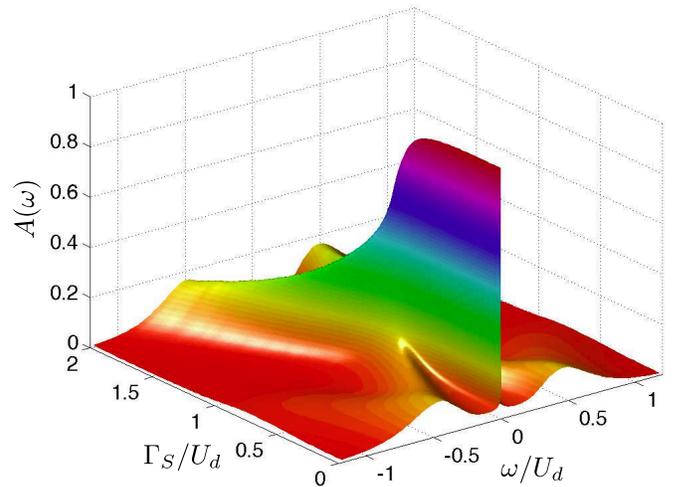}
\caption{
(Color online)
The normalized spectral function $A(\omega)$ of 
correlated quantum dot obtained by NRG calculations 
for $\Delta\gg \Gamma_{S}$ plotted
as a function of energy $\omega$ and $\Gamma_S$.
The Kondo peak is present in the doublet region, $\Gamma_S<U_d$,
while in the singlet region, $\Gamma_S > U_d$,
the Kondo peak no longer exists.
The parameters are the same as in Fig.~\ref{TK_NRG}.}
\label{TK_qcp}
\end{figure}

The estimated Kondo temperature (\ref{T_K}) increases versus $\Gamma_{S}$, 
all the way to the critical point at $\Gamma_{S}=U_{d}$. Such  tendency, 
indicated previously by the NRG data \cite{Zitko-15}, is solely caused 
by the quantum phase transition. In a  vicinity of the QPT the divergent exchange 
coupling (\ref{suppl_18}) is a typical drawback of the perturbative scheme. Figure
\ref{TK_NRG} demonstrates that the formula (\ref{T_K}) is reliable 
over the broad regime $\Gamma_{S} \leq 0.9 U_{d}$. This straightforward
conclusion can be practically used by experimentalists.

\subsection*{Equilibrium transport properties}

We now corroborate the analytical results with 
accurate numerical renormalization group calculations.
\cite{WilsonRMP75,BullaRMP08}
In NRG, the logarithmically-discretized conduction band is mapped
onto a tight binding Hamiltonian with exponentially decaying hopping,
$\xi_n \propto \Lambda^{-n/2}$, where $\Lambda$
is the discretization parameter and $n$ site index.
This Hamiltonian is diagonalized in an iterative
fashion and its eigenspectrum is then used to calculate
relevant expectation values and correlation functions.
In our calculations, we assumed $\Lambda=2$
and kept $N_k = 2048$ states during iteration
exploiting Abelian symmetry for the total spin $z$th component.~\cite{FlexibleDMNRG}
Moreover, to increase accuracy of the spectral data we averaged over
$N_z = 4$ different discretizations. \cite{oliveiraPRB05,zitkoPRB09}
We also assumed flat density of states, $\rho = 1 / 2W$,
with $W$ the band half-width used as energy unit $W \equiv 1$,
$U_d=0.1$, $\Gamma_N = 0.01$ and zero temperature.
In the absence of superconducting correlations, $\Gamma_S = 0$,
this yields the Kondo temperature, $T_K^0 \approx 10^{-5}$,
obtained from the half width at half maximum (HWHM) of the 
dot spectral function $\rho_d(\omega)$ calculated by NRG.

Figure \ref{TK_NRG}(a) presents the energy dependence
of the normalized spectral function $A(\omega)$
of the correlated quantum dot at half-filling
for the model Hamiltonian (\ref{proximized})
calculated for different values of $\Gamma_S$.
In the case of $\Gamma_S = 0$, 
$A(\omega)$ exhibits Hubbard resonance for $\omega = \pm U_d/2$
and the Kondo peak at the Fermi energy, $\omega = 0$.
It is clearly visible that increasing $\Gamma_S$ leads to the 
broadening of the Kondo peak.
In Fig.~\ref{TK_NRG}(b) we compare the relative change
of the Kondo temperature obtained from the HWHM of 
$A(\omega)$ calculated by NRG (circles)
and from the approximate formula (\ref{T_K}) based on
the generalized Schrieffer-Wolff canonical transformation (solid line).
The numerical constant $\eta$ was estimated to be $\eta = 0.6$.
The agreement is indeed very good and small deviations
occur only close to $\Gamma_S = U_d$, but then
the system is no longer in the local moment regime
and the Kondo effect disappears.

The normalized spectral function of the half-filled 
quantum dot in both the doublet, $\Gamma_S < U_d$, and
singlet region,  $\Gamma_S > U_d$, is shown in Fig. \ref{TK_qcp}.
In the doublet region we clearly observe
the zero-energy Kondo peak, whose width gradually 
increases upon increasing $\Gamma_S$.
Simultaneously the Andreev peak (whose width is roughly proportional to $\Gamma_N$)
moves toward the gap center. In the singlet state, on the other hand,
the Kondo peak does no longer exist and the Andreev peaks gradually 
depart from each other for increasing $\Gamma_{S}$. The same evolution 
of the Andreev and the Kondo quasiparticle peaks is illustrated in 
Fig.~\ref{figure2}, combining the superconducting atomic limit solution 
with the perturbative estimation of the Kondo temperature (\ref{T_K}). 


\begin{figure}
\includegraphics[width=1\columnwidth]{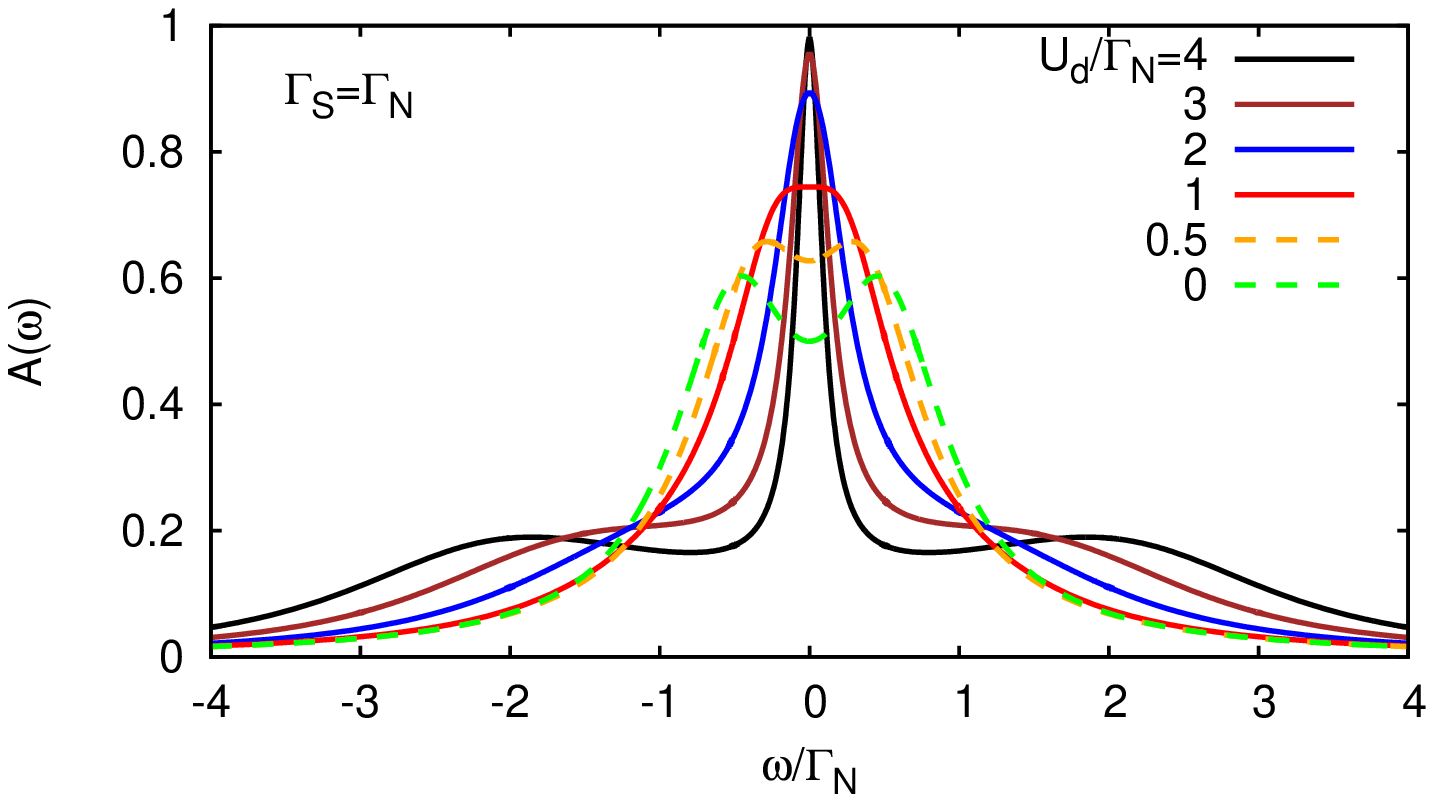}
\includegraphics[width=1\columnwidth]{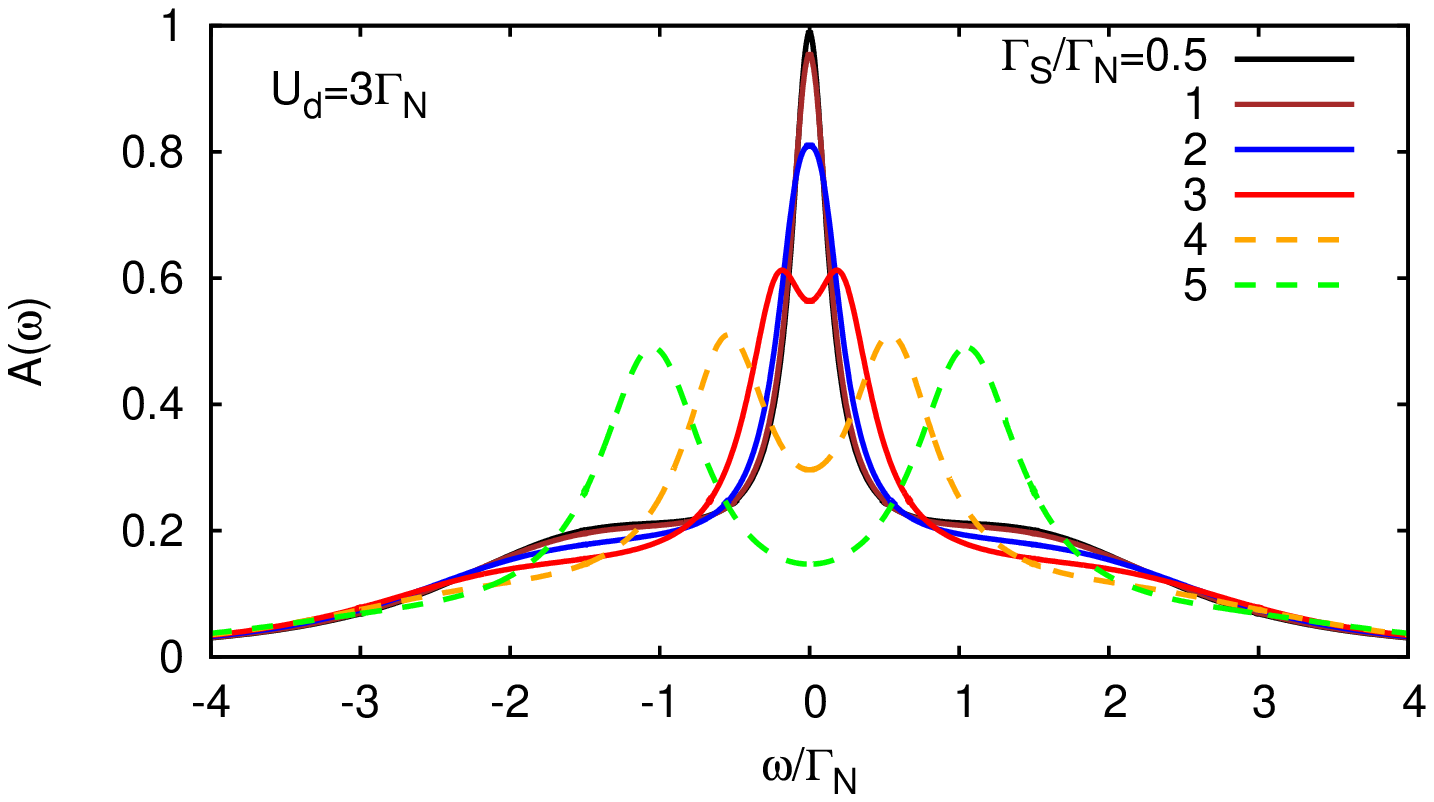}
\caption{(Color online)
The normalized spectral function $A(\omega)$ at half-filling and $T=0$
obtained from the SOPT calculations for different values of the Coulomb correlation parameter
and $\Gamma_S = \Gamma_N$ (top panel),
and for different ratios of $\Gamma_S/\Gamma_N$ with 
$U_{d} = 3\Gamma_{N}$ (bottom panel).}
\label{sopt_spectr}
\end{figure}

Broadening of the Kondo peak upon approaching the doublet-singlet transition 
can be independently supported by the second-order perturbative treatment of 
the Coulomb interaction term $U_{d} \hat{n}_{d\uparrow} \hat{n}_{d\downarrow} $. 
The first- and second-order contributions have been  discussed in 
the context of  Andreev  \cite{Cuevas-01,Yamada-11} and Josephson 
spectroscopies \cite{Vecino-2003,Meng-2009,Janis-2015}. Here we focus on 
the Kondo effect, studying its evolution near $\Gamma_{S} \sim U_{d}$.
Diagonal and off-diagonal parts of the self-energy can be expressed 
by \cite{Yamada-11} 
\begin{eqnarray}
{\mb \Sigma}_{11}(\omega) &=& -i\frac{\Gamma_{N}}{2} + U_{d}\langle
\hat{d}^{\dagger}_{\downarrow}\hat{d}_{\downarrow}\rangle 
 - \frac{U_{d}^2}{\pi} \int\limits_{-\infty}^{\infty}
{\frac{\mbox{\rm Im}{{\mb \Sigma}_{11}^{(2)}(\omega')}}
{{\omega-\omega'+i0^+}}d\omega'},
\nonumber \\ & &
\label{diag_sigma} \\
{\mb \Sigma}_{12}(\omega)&=&-\frac{\Gamma_{S}}{2} + U_{d}\langle
\hat{d}_{\downarrow}\hat{d}_{\uparrow}\rangle 
 + \frac{U_{d}^2}{\pi}\int\limits_{-\infty}^{\infty}
{\frac{\mbox{\rm Im}{{\mb \Sigma}_{12}^{(2)}(\omega')}}
{{\omega-\omega'+i0^+}}d\omega'} ,
\nonumber \\ & &
\label{offdiag_sigma} 
\end{eqnarray}
with  
\begin{widetext}
\begin{eqnarray}
\mbox{\rm Im}{\mb \Sigma}_{11(21)}^{(2)}(\omega)&=&-\int\limits_{-\infty}^{\infty}
\left[{\mb \Pi}_1(\omega+\omega')\rho_{22(21)}^+(\omega')
 + {\mb \Pi}_2(\omega+\omega')\rho _{22(21)}^-(\omega')\right]d\omega' ,
\label{imag_part}
\\
{\mb \Pi}_{1(2)}(\omega)&=&\pi\int\limits_{- \infty}^{\infty}
\left[\rho_{11}^{-(+)}(\omega')\rho_{22}^{-(+)}(\omega-\omega')
 - \rho_{12}^{-(+)}(\omega')\rho_{21}^{-(+)}(\omega-\omega') \right]\omega' . 
\label{Pi_prop}
\end{eqnarray}
\end{widetext}
In equations (\ref{imag_part},\ref{Pi_prop}) we have introduced 
$\rho_{ij}^{\pm}(\omega)=-{\pi}^{-1} \mbox{\rm Im}{\mb G}^{HF}_{ij}
(\omega)f^{\pm}(\omega)$, where $f^{\pm}(\omega)=\left[ 1 + 
\mbox{\rm exp} \left(\pm \omega/T \right)\right]^{-1}$ denotes 
the particle/hole Fermi-Dirac distribution function
and ${\mb G}^{HF}_{ij}(\omega)$ is
the Green's function obtained 
at the Hartree-Fock level 
$\hat H_{HF}  = \hat H_N  + \sum\limits_{{\bf k},\sigma } {\left( {V_{{\bf k}N} \hat d_\sigma ^{\dagger}  \hat c_{{\bf k}\sigma N}  + V_{{\bf k}N}^* \hat c_{{\bf k}\sigma N}^{\dagger}  \hat d_\sigma  } \right)}  + \sum\limits_\sigma  {\left( {\varepsilon _d  + U_d \langle \hat n_{d - \sigma } \rangle } \right)\hat n_{d\sigma}  }  - \left[ {\left( {\Delta _d  - U_d \left\langle {\hat d_ \downarrow  \hat d_ \uparrow  } \right\rangle } \right)\hat d_ \uparrow ^{\dagger}  \hat d_ \downarrow ^{\dagger}   + {\rm h}{\rm .c}{\rm .}} \right]$.
When calculating the convolutions (\ref{imag_part},\ref{Pi_prop})
we have used the  identities 
${\mb \Sigma}_{22}(\omega )=-[{\mb \Sigma}_{11}(-\omega )]^{*}$ and 
${\mb \Sigma}_{12}(\omega )=[{\mb \Sigma}_{21}(-\omega )]^{*}$.

Figure \ref{sopt_spectr} shows the spectral function $A(\omega)$ obtained
from the numerical self-consistent  solution of Eqs.\ (\ref{diag_sigma}
-\ref{Pi_prop}). For comparison with the NRG results we focused on 
the half-filled quantum dot $\langle \hat{n}_{d\sigma}\rangle=0.5$. 
In the weakly correlated case $U_{d}\leq\Gamma_{S}$ (corresponding 
to the spinless BCS-type ground state) the subgap spectrum is 
characterized by two Andreev states (shown by the dashed-line curves). 
For $U_d \sim \Gamma_S$, these Andreev states merge, forming a broad 
structure around the zero energy. In the strongly correlated 
case $U_{d}\geq\Gamma_{S}$ (corresponding to the spinfull doublet 
configuration) we observe appearance of the Kondo feature (at zero 
energy) that coexists with the  Andreev states \cite{Li_2015}. We also
notice that the width of the zero-energy peak (i.e.\ $T_{K}$) 
depends on the ratio $U_{d}/\Gamma_{S}$ and such tendency qualitatively 
agrees with our estimations based on the Schrieffer-Wolff 
transformation and with the nonperturbative NRG data.

\subsection*{Differential Andreev conductance}

\begin{figure}
\includegraphics[width=1\columnwidth]{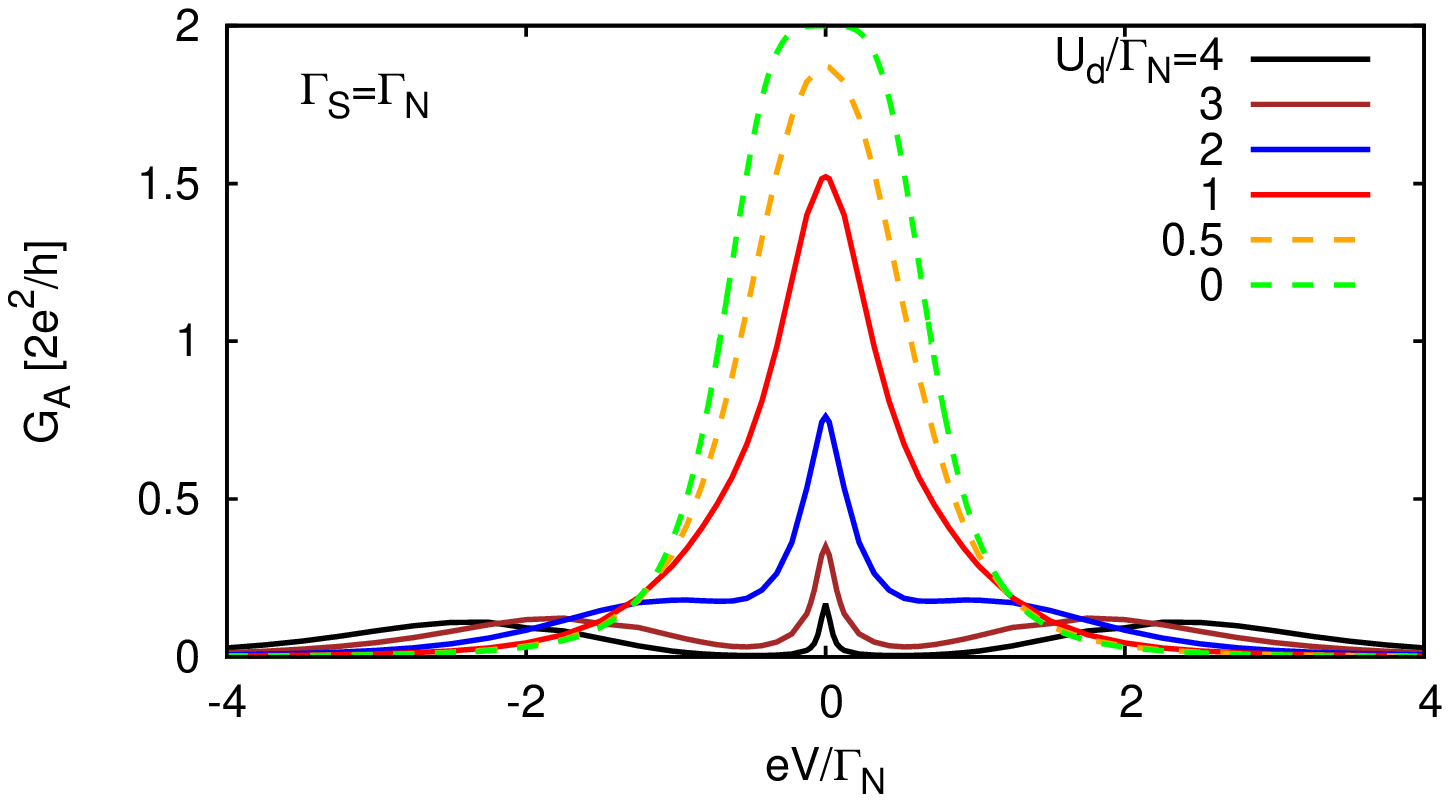}
\includegraphics[width=1\columnwidth]{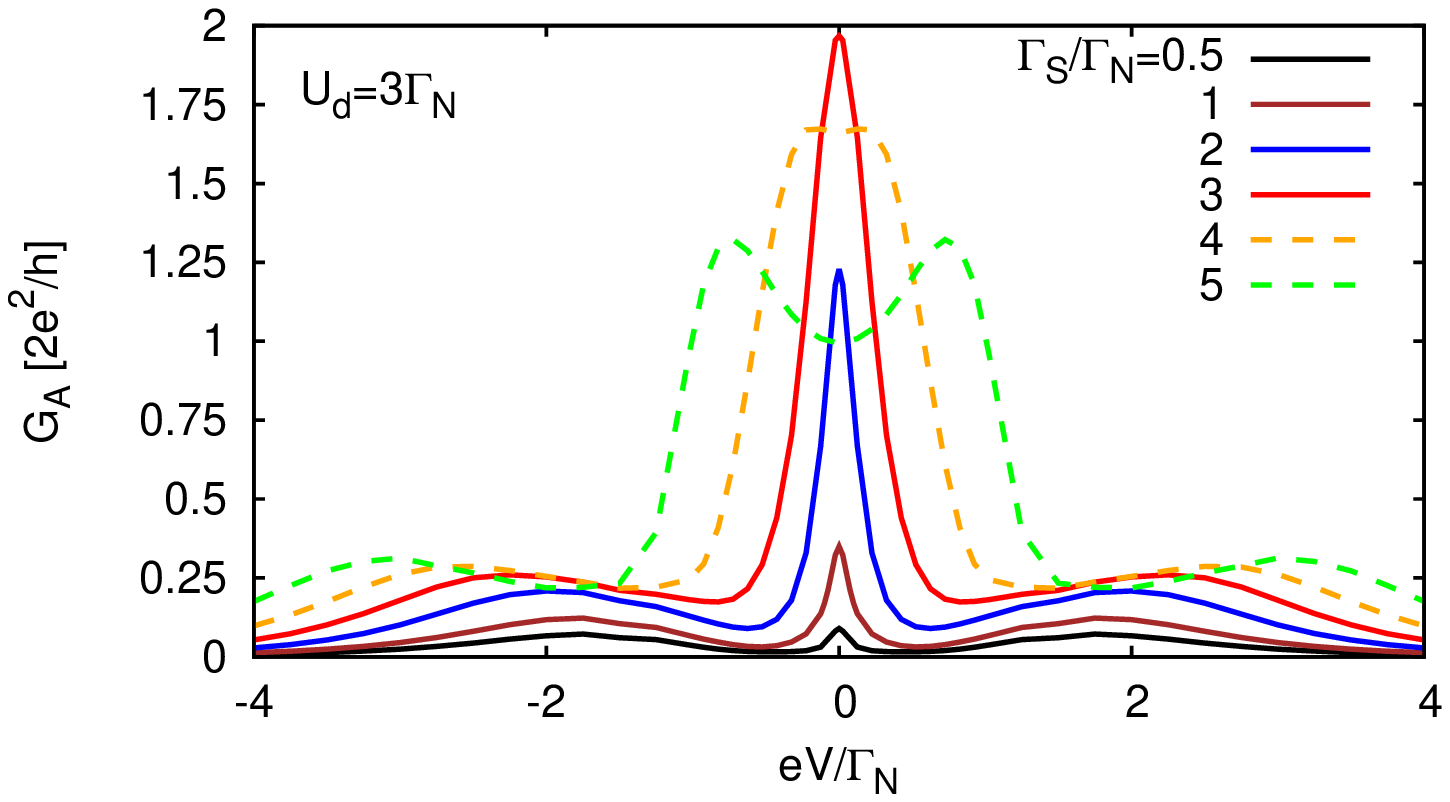}
\caption{(Color online)
The bias voltage dependence of the differential Andreev conductance $G_A$
obtained from the SOPT calculations for the 
half-filled QD at $T=0$. The left panel 
presents $G_A$  for different values of Coulomb correlation and
$\Gamma_S = \Gamma_N$, while the right panel
shows the subgap conductance 
for different values of the coupling to superconducting lead $\Gamma_S$
and for $U_d = 3\Gamma_N$.
We notice that the zero-bias feature induced 
by the Kondo effect is present only for $\Gamma_{S}<U_{d}$
(in the spinfull doublet) and its width gradually broadens with 
increasing  $\Gamma_S/U_d$.}
\label{Andr_transm}
\end{figure}

We now analyze how the observed features reveal in the 
nonlinear response regime.
For possible correspondence with the experimentally measurable quantities
we consider the subgap Andreev current 
\begin{eqnarray} 
I_{A}(V) = \frac{2e}{h} \int \!\!  d\omega \; T_{A}(\omega)
\left[ f^+(\omega\!-\!eV)\!-\!f^+(\omega\!+\!eV)\right] ,
\label{I_A}
\end{eqnarray} 
driven by the applied bias voltage $V$. The Andreev transmittance 
depends on the off-diagonal (anomalous) Green's function $T_{A}(\omega) 
= \Gamma_{N}^{2} \;  \left| {\mb G}_{12}(\omega) \right|^{2}$.
We have computed the differential conductance $G_{A}(V) = \frac{ 
\partial I_{A}(V)}{\partial V}$ determining the non-equilibrium 
transmittance by the technique described in Methods. 

Figure \ref{Andr_transm} shows the qualitative changeover of the subgap 
conductance for representative values of $U_{d}$ and $\Gamma_{S}$, corresponding to doublet
and singlet states. While approaching the QPT from the doublet side, 
we observe that the zero-bias Kondo peak is gradually enhanced, 
and its width significantly broadens. This tendency is caused by 
the characteristic Kondo temperature, which increases with 
increasing $\Gamma_{S}/U_{d}$. For $\Gamma_S > U_d$, however, the Kondo
feature is completely absent (in agreement with NRG and Schrieffer-Wolff
estimations). The magnitude of the subgap Andreev conductance approaches
then the maximum value $4e^{2}/h$ near the Andreev/Shiba states.
We notice  the quantitative difference between
the subgap transport properties 
(shown in Fig.~\ref{Andr_transm}) and the electronic spectrum
(displayed in Figs. \ref{TK_qcp} and \ref{sopt_spectr}).
Observability of the Kondo enhancement would be thus possible
only close to the QPT on the doublet side.

\section*{Discussion}

We have studied the influence of the electron pairing on 
the Kondo effect in the strongly correlated quantum dot coupled 
(by $\Gamma_{N}$) to the metallic  and  (by $\Gamma_{S}$) to 
superconducting reservoirs by three independent methods. 
The proximity induced on-dot pairing and the Coulomb repulsion 
$U_{d}$ are responsible for the quantum 
phase transition between the (spinless) BCS-like singlet and 
the (spinfull) doublet configurations, depending on the ratio 
of $\Gamma_S/U_d$. Upon approaching this quantum critical point
from the doublet side, one observes the enhancement of 
the Kondo temperature with increasing $\Gamma_S$ \cite{Zitko-15}.
We have provided the microscopic arguments supporting this behavior 
based on the generalized Schrieffer-Wolff canonical transformation.
This perturbative treatment of the coupling to metallic lead
revealed enhancement of the antiferromagnetic spin-exchange 
potential, responsible for the Kondo resonance. 
We have compared the estimated Kondo temperature with the numerical 
renormalization group calculations, and found excellent agreement 
over the broad regime $\Gamma_{S}<0.9U_{d}$. We have confirmed 
this tendency (for arbitrary $\Gamma_{N}$) using the second-order 
perturbative treatment of the Coulomb interaction. Our 
analytical estimation of the Kondo temperature (\ref{T_K})
can be quantitatively verified in experimental measurements
of the differential Andreev conductance. We have shown, 
that the zero-bias enhancement of the subgap conductance 
(already reported \cite{Deacon-10,Aguado-13,Lohneysen-12,Chang-13}
for some fixed values of $\Gamma_{S}$) would be significantly
amplified with increasing ratio $\Gamma_{S}/U_d$, but only on
the doublet side. Such behaviour is in stark contrast with
the zero-bias anomaly caused by the Majorana quasiparticles
resulting from the topologically non-trivial superconductivity.

\section*{Appendices}

{\footnotesize

\subsection*{The deep subgap regime $|\omega| \ll \Delta$}
\label{A}

When studying the proximity effect of the Anderson-type Hamiltonian 
(\ref{model}) one has to consider the mixed particle and hole degrees 
of freedom. This can be done, by defining the matrix Green's function 
${\mb G}(\tau,\tau') \!=\! \langle\!\langle \hat{\Psi}_{d}(\tau); 
\hat{\Psi}_{d}^{\dagger}(\tau')\rangle\!\rangle$ in the Nambu spinor
representation, $\hat{\Psi}_{d}^{\dagger}=(\hat{d}_{\uparrow}^{\dagger},
\hat{d}_{\downarrow})$, $\hat{\Psi}_{d}=(\hat{\Psi}_{d}^{\dagger})
^{\dagger}$. Here we determine its diagonal and 
off-diagonal parts in the equilibrium case (which is also useful 
for description of the transport  within the Landauer formalism). 
The Fourier transform of the Green's function ${\mb G}_{d}(\tau,\tau')
={\mb G}_{d}(\tau-\tau')$ can be expressed by the Dyson equation
\begin{eqnarray} 
{\mb G}^{-1}(\omega) = 
\left( \begin{array}{cc}  
\omega\!-\!\varepsilon_{d} &  0 \\ 0 &  
\omega\!+\!\varepsilon_{d}\end{array}\right)
- {\mb \Sigma}_{d}(\omega) . 
\label{GF}\end{eqnarray} 
The self-energy ${\mb \Sigma}_{d}(\omega)$ accounts for the coupling 
of the quantum dot to external reservoirs and for the correlation
effects originating from the Coulomb repulsion $U_{d}$.

The quantum dot hybridization with the leads
can be expressed analytically by
${\mb \Sigma}_{d}^{(U_{d}\!=\!0)}(\omega)=\sum_{{\bf k},{\beta}}|V_{{\bf k}
\beta}|^{2}g_{{\bf k}\beta}(\omega)$, where $g_{{\bf k}\beta}(\omega)$ 
are the (Nambu) Greens' functions of itinerant electrons. 
In the wide-band limit this self-energy is given by the following 
explicit formula \cite{Bauer-07,Yamada-11} 
\begin{eqnarray}
\mb{\Sigma}_{d}^{(U_{d}\!=\!0)}(\omega) &=&  
- \; \frac{i \Gamma_{N}}{2} \; \left( \begin{array}{cc}  
1 & 0 \\ 0 & 1 \end{array} \right)
 -  \frac{\Gamma_{S}}{2} \left( \begin{array}{cc}  
1 & \frac{\Delta}{\omega} \\ 
 \frac{\Delta}{\omega}  & 1 
\end{array} \right) 
\nonumber \\ & & \times 
\left\{
\begin{array}{ll} 
\frac{\omega}{\sqrt{\Delta^{2}-\omega^{2}}}
& \mbox{\rm for }  |\omega| < \Delta  \\
\frac{i\;|\omega|}{\sqrt{\omega^{2}-\Delta^{2}}}
& \mbox{\rm for }  |\omega| > \Delta 
\end{array} \right. . 
\label{selfenergy_0}
\end{eqnarray} 
Equation (\ref{selfenergy_0}) describes: (i) the proximity 
induced on-dot pairing (via the term proportional to $\Gamma_{S}$) 
and (ii) the broadening (finite life-time) effects. The latter 
come from the imaginary parts of self-energy (\ref{selfenergy_0})
and depend either on both couplings $\Gamma_{\beta =N,S}$ 
(for energies $|\omega|\geq\Delta$) or solely on $\Gamma_{N}$ 
(in the subgap regime  $|\omega|<\Delta$).

In the subgap regime $|\omega|<\Delta$ the Green's function 
of uncorrelated quantum dot acquires the BCS-type structure
\begin{eqnarray}
{\mb G}(\omega) \!=\!
\left( \! \begin{array}{cc}  
\tilde{\omega} + i \Gamma_{N}/2 - \varepsilon_{d}   
\hspace{0.3cm}&  \tilde{\Gamma}_{S}/2 \\ 
\tilde{\Gamma}_{S}/2  & 
\tilde{\omega} + i \Gamma_{N}/2 + \varepsilon_{d}   
\end{array} \! \right)^{\!\!-1} 
\label{G_0}
\end{eqnarray}
with  $\tilde{\omega} = \omega + \frac{\Gamma_{S}}{2} \frac{\omega}
{\sqrt{\Delta^{2}-\omega^{2}}}$ and $\tilde{\Gamma}_{S} = \Gamma_{S} 
\frac{\Delta}{\sqrt{\Delta^{2}-\omega^{2}}}$. The resulting spectrum 
consists of two in-gap peaks, known as the Andreev \cite{Bauer-07,Andreev} 
or Yu-Shiba-Rusinov \cite{Yu-Shiba-Rusinov,BalatskyRMP06} quasiparticles. Their 
splitting is a measure of the pairing gap $\Delta_{d}$ induced in the quantum dot.
Figure \ref{figure1} displays the characteristic energy scales 
of the uncorrelated quantum dot.

\begin{figure}
\includegraphics[width=1\columnwidth]{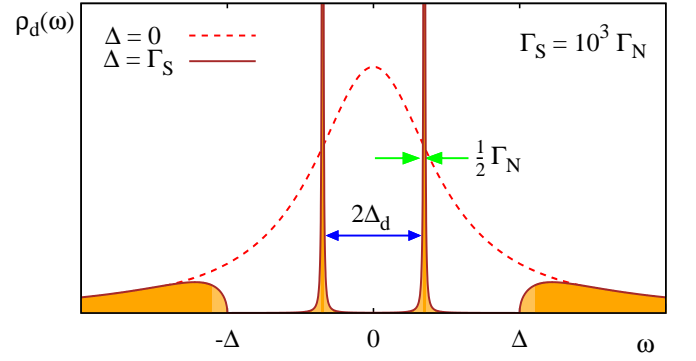}
\caption{(Color online)Spectral function $A(\omega)$ of the uncorrelated QD
 obtained for $\varepsilon_{d}=0$, $\Gamma_{N} = 10^{-3}$
and $\Delta = \Gamma_{S}$.
The dashed line shows the reference spectrum 
in the absence of superconducting correlations, $\Delta=0$. 
The in-gap states are separated by $2\Delta_{d}$.}
\label{figure1}
\end{figure}

For infinitesimally weak coupling $\Gamma_{N}\!=\!0^{+}$ the in-gap 
states have a shape of Dirac delta functions (i.e.\ represent the long-lived 
quasiparticles). Otherwise, they acquire a finite broadening proportional 
to $\Gamma_{N}$. In the absence of correlations (for $U_{d}\!=\!0$) 
the quasiparticle energies $E_{A,\pm}$ can be determined by solving 
the following equation \cite{Domanski-08,Baranski-13}
\begin{eqnarray}
E_{A,\pm} +  \frac{(\Gamma_{S}/2)E_{A,\pm}} {\sqrt{\Delta^{2}
-E_{A,\pm}^{2}}} = \pm \sqrt{\varepsilon_{d}^{2}+ 
\frac{(\Gamma_{S}/2)^{2}\Delta^{2}}{\Delta^{2}-E_{A,\pm}^{2}}} .
\label{energy_eqn}
\end{eqnarray}
In the strong coupling limit, $\Gamma_{S}\gg \Delta$, we can notice
that in-gap quasiparticles appear close to the superconductor gap
edges $E_{A,\pm}\simeq \pm\Delta$, whereas in the weak coupling 
limit, $\Gamma_{S}\ll \Delta$, they approach the asymptotic values, 
$E_{A,\pm} \simeq \pm \sqrt{\varepsilon_{d}+\left(\Gamma_{S}/2
\right)^{2}}$. For $\Gamma_{N}\rightarrow 0$, the latter case is 
known as the 'superconducting atomic limit'. The self-energy 
(\ref{selfenergy_0}) simplifies then to the static value
\begin{eqnarray}
\mb{\Sigma}_{d}^{0}(\omega) &=&  - \; \frac{1}{2} \; 
\left( \begin{array}{cc}  i\Gamma_{N} & \Gamma_{S} \\ 
\Gamma_{S} & i\Gamma_{N} \end{array} \right) ,
\end{eqnarray}
therefore the  Hamiltonian (\ref{model}) can be formally 
modeled by its fully equivalent form (\ref{proximized}), 
describing the proximized quantum dot coupled to the metallic lead. 

\subsection*{Influence of the correlation effects}

We note that the early studies of the nontrivial relationship between the Coulomb repulsion 
and the proximity induced electron pairing of the normal metal - quantum dot 
- superconductor (N-QD-S) junctions have adopted variety of the theoretical 
methods, such as: slave boson approach \cite{Fazio-98,Kang-98}, equation 
of motion \cite{Sun-99}, noncrossing approximation \cite{Clerk-00}, iterative
perturbation technique \cite{Cuevas-01}, path integral formulation of the 
dynamical mean field approximation \cite{Avishai-01}, constrained slave
boson method \cite{Krawiec-04}, numerical renormalization group 
\cite{Bauer-07,Weymann-14,Zitko-15,Tanaka-07}, modified equation of motion 
\cite {Domanski-08}, functional renormalization group \cite{Karrasch-08}, 
expansion around the superconducting atomic limit \cite{Meng-10}, 
cotunneling treatment of the spinful dot \cite{Paaske-10}, numerical 
QMC simulations \cite{Koga-13}, selfconsistent perturbative
treatment of the Coulomb repulsion \cite{Yamada-11} and other
\cite{Konig-10,Rodero-11,Baranski-13}. Amongst them  only the  
numerical renormalization group (NRG) calculations \cite{Zitko-15} 
suggested the Kondo temperature to exponetially increase with
increasing $\Gamma_{S}$ when approaching the quantum phase transition 
from the doublet side ($\Gamma_S \sim U_{d}$). 

The relationship between the proximity induced on-dot pairing and the screening 
effects can be better understood by analyzing the superconducting order 
parameter $\langle \hat{d}_{\downarrow} \hat{d}_{\uparrow} \rangle$ and
the QD magnetization $\langle \hat{S}_{d}^{z} \rangle = \frac{1}{2}
\expect{ \hat{d}^\dag_\uparrow \hat{d}_\uparrow - \hat{d}^\dag_\downarrow 
\hat{d}_\downarrow}$. In \fig{proximity_effect} we show their
dependence on the coupling $\Gamma_S$ for several $\Gamma_{N}/U_{d}$ ratios
calculated by NRG.
For finite superconducting energy gap
a sign change of the order parameter signals the 
quantum phase transition
\cite{Zitko-15}. However, in the case of infinite gap considered here,
$\langle \hat{d}_{\downarrow} \hat{d}_{\uparrow} \rangle$
only drops to zero at the transition point
\cite{Bauer-07,Tanaka-07}.
As clearly seen in the figure, 
the order parameter  $\langle 
\hat{d}_{\downarrow} \hat{d}_{\uparrow} \rangle$ increases from $0$ to $\frac{1}{2}$
around $\Gamma_S \sim U_{d}$ [\fig{proximity_effect}(a)]
corresponding to the QPT. Its enhancement is accompanied by 
the suppression of the dot magnetization, which vanishes in 
the singlet phase, $\Gamma_S > U_d$, $\langle \hat{S}_{d}^{z} \rangle \to 0$,
see \fig{proximity_effect}(b).
Both the increase of $\langle \hat{d}_{\downarrow} \hat{d}_{\uparrow} \rangle$
and the decrease of $\langle \hat{S}_{d}^{z} \rangle$ indicate
the quantum phase transition at $\Gamma_S = U_d$.
Moreover, it can be also seen that the transitions present
in the above quantities become smeared with increasing
the coupling to the normal lead $\Gamma_N$.

\begin{figure}
\includegraphics[width=1\columnwidth]{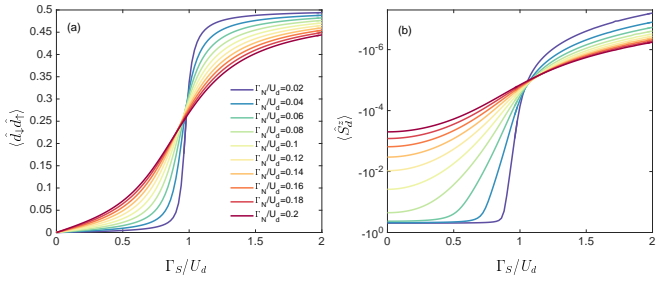}
\caption{(Color online)(a) The superconducting order parameter
$\langle \hat{d}_{\downarrow}  \hat{d}_{\uparrow}\rangle$
and (b) the magnetization $\langle \hat{S}_{d}^{z} \rangle$
of the correlated quantum dot calculated by NRG
for different coupling to normal lead $\Gamma_N$, as indicated.
The parameters are the same as in Fig.~\ref{TK_NRG}.
In panel (b) a small external magnetic field $B$ is applied
to the system, $B = 10^{-6}$.}
\label{proximity_effect}
\end{figure}

\subsection*{Nonlinear charge transport in the subgap regime}

Under non-equilibrium conditions the Andreev transmittance
$T_A (\omega ) = \Gamma _N^2 \left| {G_{12} (\omega )} \right|^2$ 
has to be determined using the lesser and greater self-energies 
\cite{Yamada-11}
\begin{eqnarray}
\Sigma^{<,>}(\omega ) & = & \Sigma^{<,>0}(\omega ) + \Sigma^{<,> 2}(\omega ) 
\nonumber \\
\Sigma ^{ < 0}(\omega )  & = & i\left( {\begin{array}{*{20}c}
   {\Gamma _N f^ +  (\omega - eV)} & 0  \\
   0 & {\Gamma _N f^ +  (\omega + eV )}  \\
\end{array}} \right)
\nonumber \\
\Sigma^{ > 0}(\omega )  & = &  - i\left( {\begin{array}{*{20}c}
   {\Gamma _N f^ -  (\omega - eV)} & 0  \\
   0 & {\Gamma _N f^ -  (\omega + eV)}  \\
\end{array}} \right)
\nonumber
\end{eqnarray}
where
\begin{eqnarray}
\Sigma _{11}^{ <,> 2} (\omega) &=& \frac{{U_d^2}}{{2\pi }}
\int\limits_{ - \infty }^\infty  {\Pi^{ <,> } (\omega' + 
\omega )\rho _{22}^{ >,< } (\omega')} d\omega' 
\nonumber \\
\Sigma _{22}^{ <,> 2} (\omega) &=& \frac{{U_d^2}}{{2\pi }}
\int\limits_{ - \infty }^\infty  {\Pi^{ <,> } (\omega ' 
+ \omega )\rho _{11}^{ >,< } (\omega')} d\omega'
\nonumber \\
\Sigma _{12}^{ < , > 2} (\omega ) &=&  - \frac{{U_d^2 }}{{2\pi }}
\int\limits_{ - \infty }^\infty  {\Pi ^{ < , > } (\omega ' 
+ \omega )\rho _{12}^{ > , < } (\omega ')} d\omega '
\nonumber \\
\Pi ^{ < , > } (\omega ) &=& \frac{1}{{2\pi }}
\int\limits_{ - \infty }^\infty  {\left[ {\rho _{11}^{ < , > } 
(\omega ')\rho _{22}^{ < , > } (\omega  - \omega ') 
- \rho _{12}^{ < , > } (\omega ')\rho _{21}^{ < , > } 
(\omega  - \omega ')} \right]} d\omega '
\nonumber
\end{eqnarray}
and $\rho^{ < , > }  = G^{HF,r} \Sigma ^{ < , > 0} G^{HF,a}$,
with $G^{HF,r(a)}$ denoting the respective retarded (advanced)
Green's function.
The expectation values 
$\left\langle {\hat d_\sigma ^ \dag  \hat d_\sigma  } \right\rangle$ 
and $\left\langle {\hat d_ \downarrow  \hat d_ \uparrow  } \right\rangle$ 
have been determined self-consistently from  $\left\langle {\hat d_\sigma ^\dag  \hat d_\sigma  } 
\right\rangle  = \frac{1}{{2\pi i}}\int\limits_{ - \infty }^\infty  {G_{11}^ <  
\left( \omega  \right)d\omega } $ and $\left\langle {\hat d_ \downarrow  
\hat d_ \uparrow  } \right\rangle  = \frac{1}{{2\pi i}}
\int\limits_{ - \infty }^\infty  {G_{12}^ <  \left( \omega  \right)d\omega }$, 
where the lesser and greater Greens' functions obey $G^{ <,> }(\omega)  = 
G^r(\omega) \Sigma ^{ <,> }(\omega) G^a(\omega)$, as discussed
in Ref.\ \cite{Yamada-11}.

} 


\section*{Acknowledgments}

T.D.\ kindly acknowledges R.\ \v{Z}itko for  inspiring discussions 
and thanks K.\ Franke and T.\ Meng for valuable remarks.
This work is supported by the National Science Centre in Poland 
through the projects DEC-2014/13/B/ST3/04451 (TD) and
DEC-2013/10/E/ST3/00213 (IW), and
the Faculty of Mathematics and Natural Sciences 
of the University of Rzesz\'ow through the project WMP/GD-06/2015 (GG).

\section*{Author contributions statement}

T.D. arranged the project, M.Z. constructed the Schrieffer-Wolff transformation,
I.W. performed the NRG calculations, and G.G.\ solved the SOPT
equations. All authors consulted the results. The manuscript 
was prepared by T.D. and I.W.

\section*{Additional information}
The authors declare no competing financial interests.

\end{document}